\documentclass[aps,showpacs,floatfix,twocolumn]{revtex4}
\usepackage{graphics}
\usepackage{dcolumn}
\usepackage{amsmath}
\usepackage{booktabs}
\usepackage{amssymb}
\usepackage{graphics,subfig}
\usepackage{rotating}
\usepackage{bm}
\usepackage{float}
\usepackage[T1]{fontenc}
\newcommand{\bea}{\begin{eqnarray}}
\newcommand{\eea}{\end{eqnarray}}
\newcommand{\be}{\begin{equation}}
\newcommand{\ee}{\end{equation}}
\newcommand{\f}[1]{\frac{1}{2}}
\def\be{\begin{eqnarray}}
\def\ee{\end{eqnarray}}
\def\bd{\begin{displaymath}}
\def\ed{\end{displaymath}}
\def\nn{\nonumber}

\def\etal{{\em et al. }}

\def\NP{Nucl. Phys. }
\def\PR{Phys. Rev. }

\def\epl{Eur. Phys. Lett.}

\begin{document}
\title{Massive neutron stars with hyperonic core : a case study with the IUFSU model}
\author{Bipasha Bhowmick}
\email{bips.gini@gmail.com}
\author{Madhubrata Bhattacharya}
\email{madhubrata.b@rediffmail.com}
\author{Abhijit Bhattacharyya}
\email{abphy@caluniv.ac.in}
\author{ G. Gangopadhyay}
\email{ggphy@caluniv.ac.in}
\affiliation{Department of Physics, University of Calcutta,\\
92 Acharya Prafulla Chandra Road, Kolkata-700 009, India}
\begin{abstract}
The recent discoveries of massive neutron stars, such as PSR J$0348+0432$ and PSR J$1614-2230$, have raised 
questions about the existence of exotic matter such as hyperons in the neutron star core. The validity of many 
established equations of states (EoS's) like the GM1 and FSUGold are also questioned. We investigate the existence 
of hyperonic matter in the central regions of massive neutron stars using Relativistic Mean Field (RMF) theory 
with the recently proposed IUFSU model. The IUFSU model is extended by including hyperons to study the 
neutron star in $\beta$ equilibrium. 
The effect of different hyperonic potentials, namely $\Sigma$ and $\Xi$ potentials, on the EoS and hence 
the maximum mass of neutron stars has been studied. We have also considered the effect of stellar rotation since the observed 
massive stars are pulsars.
 It has been found that a maximum mass of 
$1.93M_{\odot}$, which is within the 3$\sigma$ limit of the observed mass of PSR J$0348+0432$, can be obtained for 
rotating stars, with certain choices of the hyperonic potentials. The said star
 contains a fair amount of hyperons near the core.
\vskip 0.2cm
\end{abstract}
\maketitle
\section{Introduction}

The recent discoveries of the massive neutron stars PSR J$0348+0432$~\cite{science} and PSR J$1614-2230$~\cite{Nature} have 
brought new challenges for theories of dense matter
beyond the nuclear saturation density. Recently the radio timing measurements of the pulsar PSR J$0348+0432$ and its white dwarf 
companion have confirmed the mass of the pulsar to be in the
range of $1.97 - 2.05$ M$_{\odot}$ at $68.27\%$ or $1.90 - 2.18$ M$_{\odot}$ at $99.73\%$ confidence~\cite{science}. This 
is only the second neutron star(NS) with a precisely determined mass around 2M$_{\odot}$, after PSR J$1614-2230$
and has a 3$\sigma$ lower mass limit $0.05$ M$_{\odot}$ higher than the latter. It therefore
 provides the tightest reliable lower bound on the maximum mass of neutron
stars.

Compact stars provide the perfect astrophysical environment for testing theories of cold 
and dense matter. Densities at the core of neutron stars can reach values of several times of
$10^{15} gm \,\, cm^{-3}$. At such high densities, the energies of the particles are high enough to favour the appearance of exotic particles
in the core. Since the lifetime of neutron stars are much
greater than those associated with the weak interaction, strangeness
conservation can be violated in the core due to the weak interaction. This would result in the appearance
of strange particles such as hyperons. The appearance of such particles produces new degrees of freedom, which results in a
softer equation of state (EoS) in the neutron star interior. 

The observable properties of compact stars depend crucially on the EoS.
According to the existing models of dense matter the presence of strangeness in the neutron star interior
leads to a considerable softening of the EoS, resulting in a reduction of
the maximum mass of the neutron star~\cite{2,3,4,5}.
Therefore many existing theories involving hyperons cannot explain
the large pulsar masses~\cite{6}. Most relativistic models obtain maximum neutron
star masses in the range $1.4-1.8M_{\odot}$~\cite{7,8,9,10,11,12,13,14}, when hyperons are included.
Some authors have tackled this problem by including a strong vector repulsion in the strange sector or by pushing 
 the threshold for the appearance of 
hyperons to higher densities~\cite{14,15,16,17,18,19,20,21}.
In
several studies the maximum neutron star masses were generally
found to be lower than $1.6M_\odot$~\cite{3,4,5,22,23,24,25,26} which is in contradiction with observed pulsar masses. 
However, neutron stars with maximum mass
larger than $2M_{\odot}$ have been obtained theoretically. 
 Bednarek \etal~\cite{27} achieved a stiffening 
of the EoS by using a non-linear relativistic mean
field (RMF) model with quartic terms involving the strange vector meson. 
 Lastowiecki \etal~\cite{28} obtained massive stars
 including a quark matter core. Taurines \etal~\cite{29} achieved large neutron star masses including
hyperons by considering a model with density dependent coupling constants. The coupling constants were varied nonlinearly with the
scalar field. Bonanno and Sedrakian~\cite{30} also modeled massive neutron stars including hyperons and quark
core using a fairly stiff EoS and vector repulsion among quarks.
Authors in ref.~\cite{Gupta} incorporated higher order couplings in the RMF theory in addition to kaonic interactions 
to obtain the maximum neutron star mass. Agrawal \etal\cite{Agrawal1} have optimized the parameters of the extended RMF model using a selected set of
global observables which includes binding energies and charge radii for nuclei along several isotopic
and isotonic chains and the iso-scalar giant monopole resonance energies for the $^{90}$Zr and $^{208}$Pb
nuclei. 
Weissenborn \etal~\cite{30a} investigated the vector meson-hyperon coupling, 
going from SU(6) quark model to a broader SU(3), and concluded that the maximum mass of a neutron star decreases 
linearly with the strangeness content of the
neutron star core independent of the nuclear EoS.
On the other hand, H. Dapo \etal~\cite{5} found that for several different bare
hyperon-nucleon potentials and a wide range of nuclear matter parameters the hyperons in neutron
stars are always present.

 The parameters of the RMF model are fitted to the saturation properties of the infinite nuclear matter and/or the properties 
of finite nuclei. 
As a result extrapolation to higher densities and asymmetry involve uncertainties. Three of these properties of the 
infinite nuclear matter are more precisely known: (a) the saturation density,
(b) the binding energy and (c) the asymmetry energy, compared to
the remaining ones - the effective nucleon mass and the compression modulus of the nuclear matter. 
The uncertainty in the dense matter EoS is basically
related to the uncertainty in these two saturation properties. It has been seen that to reproduce the giant monopole resonance (GMR)
 in $^{208}$Pb, accurately fitted non-relativistic and relativistic models predict compression
modulus in the symmetric nuclear matter ($K$) that differ by
about $25\%$. The reason for this discrepancy being the density dependence of the symmetry energy. 
Moreover, the alluded correlation between $K$ and the density dependence of the symmetry energy 
results in an underestimation of the frequency of oscillations of
neutrons against protons, the so-called isovector giant
dipole resonance (IVGDR) in $^{208}$Pb. FSUGold is a recently proposed accurately calibrated relativistic 
parameterization. It simultaneously describes
the GMR in $^{90}$Zr and $^{208}$Pb and the IVGDR in $^{208}$Pb without compromising the success in reproducing the ground-state
observables~\cite{30b}. The main virtue of this parameterization is the softening of both the EoS of symmetric nuclear matter and the symmetry energy. 
This softening appears to be required for an
accurate description of different collective modes having different neutron-to-proton ratios.
As a result, the FSUGold effective interaction predicts neutron star radii that are too large and a maximum stellar
mass that is too small~\cite{31}. 

 The Indiana University-Florida State
University (IUFSU) interaction, is a new relativistic parameter set, derived from FSUGold. It is simultaneously constrained by
the properties of finite nuclei, their collective excitations and the neutron star properties by adjusting
two of the parameters of the theory - the neutron skin thickness
of $^{208}Pb$ and the maximum neutron star mass~\cite{32}. As a result the new effective interaction
softens the EoS at intermediate densities and stiffens the
EoS at high density. As it stands now, the
new IUFSU interaction reproduces 
the binding energies and charge radii of closed-shell
nuclei, various nuclear giant (monopole and dipole)
resonances, the low-density behavior of pure neutron
matter, the high-density behavior of the symmetric nuclear
matter and the mass-radius relationship of neutron stars. Whether this new EoS can accommodate 
the hyperons inside the compact stars, with the severe constraints imposed by the recent observations of $\sim 2M_\odot$ pulsars,
needs to be explored. In this work we plan to make a detailed study of such a possibility. For this purpose
we have extended the IUFSU interaction by including 
the full baryon octet.  A new EoS is constructed to investigate the neutron star properties with  
hyperons.

\begin{table*}[t] 

\begin{tabular}{|c|c|c|c|c|c|c|c|c|}\hline

Model &$g_{\sigma n}^2$ & $g_{\omega n}^2$& $g_{\rho n}^2$ & $\kappa$  &   $\lambda$   &   $\zeta$   &   $\Lambda_v $\\[0.1cm]
&          &              &              &                 (MeV)          &                  &              &    \\\hline
FSU   &   112.1996   &   204.5469   &   138.4701   & 1.4203   &   0.023762   &   0.06   &   0.030\\[0.1cm]
IUFSU &   99.4266   &   169.8349   &   184.6877    & 3.3808   &   0.000296   &   0.03   &   0.046\\[0.1cm]
\hline
\end{tabular}

\caption{ Parameter sets for the two models discussed in the text. 
The nucleon mass and the meson masses are kept fixed at $m_n$ = 939 MeV, $m_\sigma$ = 491.5 MeV,
$m_\omega$ = 782.5 MeV, $m_\rho$ = 763 MeV and $m_\phi$ = 1020 MeV in both the models.
\label{Para}}

\end{table*}

The paper is organized as follows. In section 2, we briefly discuss the
model used and the resulting EoS. In the next section we use this EoS 
to look at static and rotating star properties. We give a brief summary in section 4.

\section{ IUFSU with hyperons}
 One of the possible approaches to describe neutron star matter is to adopt
an RMF model subject to $\beta$ equilibrium and charge neutrality. For our
investigation of nucleons and hyperons in the compact star matter we choose
the full standard baryon octet as well as electrons and muons. Contribution from neutrinos
are not taken into account assuming that they can escape freely from the system. In
this model, baryon-baryon interaction is mediated by the exchange of scalar
($\sigma$), vector ($\omega$), isovector ($\rho$) and the strange vector ($\phi$) mesons. 
The Lagrangian density we consider is given by~\cite{32}
\begin{widetext}
\begin{eqnarray}
\mathcal{L} &=& \sum_{B}\bar{\psi}_{B}[i\gamma^{\mu}\partial_{\mu}
- m_{B}+g_{\sigma B}\sigma - g_{\omega B}\gamma^{\mu}\omega_{\mu}  - g_{\phi B}\gamma^{\mu}\phi_{\mu}
- \frac{g_{\rho B}}{2}\gamma^{\mu}\vec{\tau}\cdot\vec{\rho}^{\mu}]{\psi}_{B} +
\frac{1}{2}\partial_{\mu}\sigma\partial^{\mu}\sigma 
- \frac{1}{2} m_\sigma^2\sigma^2 \nn\\
&& - \frac{\kappa}{3!}(g_{\sigma N}\sigma)^3  -\frac{\lambda}{4!}(g_{\sigma N}\sigma)^4 - \frac{1}{4}F_{\mu\nu}F^{\mu\nu} +
\frac{1}{2}m_\omega^2\omega_\mu\omega^\mu  + \frac{\zeta}{4!}(g^{2}_{\omega N}\omega_\mu\omega^\mu)^2 
+\frac{1}{2}m_\rho^2\vec{\rho}_{\mu}\cdot\vec{\rho}^{\mu} - \frac{1}{4}\vec{G}_{\mu\nu}\vec{G}^{\mu\nu}\nn\\ 
&&+ \Lambda_{v}(g^{2}_{\rho N}\vec{\rho}_{\mu}\cdot\vec{\rho}^{\mu})(g^{2}_{\omega N}\omega_\mu\omega^\mu)  
+ \frac{1}{2}m_\phi^2\phi_\mu\phi^\mu -\frac{1}{4}H_{\mu\nu}H^{\mu\nu} 
+\sum_{l}\bar{\psi}_{l}[i\gamma^{\mu}\partial_{\mu} - m_{l}]{\psi}_{l}
\end{eqnarray}
\end{widetext}

\noindent where the symbol B stands for the baryon octet ($p$, $n$, $\Lambda$, $\Sigma^{+}$, $\Sigma^{0}$, $\Sigma^{-}$,
$\Xi^{-}$, $\Xi^{0}$) and $l$ represents $e^{-}$ and $\mu^{-}$. The masses $m_B$, $m_\sigma$, 
$m_\omega$, $m_\rho$ and $m_\phi$ are respectively for baryon, $\sigma$, $\omega$, $\rho$ and $\phi$ mesons. The antisymmetric 
tensors of vector mesons take the forms ${F}
_{\mu\nu}$ =  $\partial_{\mu}\omega_{\nu} - \partial_{\nu}\omega_{\mu}$,
${G}_{\mu\nu}$ =  $\partial_{\mu}\vec{\rho}_{\nu} - \partial_{\nu}\vec{\rho}_{\mu} + g[\vec{\rho}_{\mu},{\vec\rho}_{\nu}]$ and
${H}_{\mu\nu}$ =  $\partial_{\mu}{\phi}_{\nu} - \partial_{\nu}{\phi}_{\mu}$. The isoscalar meson 
self-interactions (via $\kappa$, $\lambda$ and $\zeta$ terms) are necessary for the appropriate EoS of the
symmetric nuclear matter~\cite {33}. The new additional isoscalar-isovector coupling ($\Lambda_v$) term 
is used to modify the density dependence of the symmetry energy and the neutron-skin thickness of heavy 
nuclei~\cite {31,32}. The meson-baryon coupling constants are given by
 $g_{\sigma B}$, $g_{\omega B}$, $g_{\rho B}$ and $g_{\phi B}$.

All the nucleon-meson parameters used in this work are shown in Table  \ref{Para}. The saturation properties of the symmetric 
nuclear matter produced by IUFSU are: saturation density $n_0=0.155$ $fm^{-3}$, binding energy per nucleon  
$\varepsilon_0= -16.40$ MeV and compression modulus $K = 231.2$ MeV.

The hyperon-meson couplings are taken from the SU(6) quark model~\cite{34,35} as,
\begin{center}
$g_{\rho\Lambda}$ = 0, $g_{\rho\Sigma}$ = $2g_{\rho\Xi}$ = $2g_{\rho N}$
\end{center}
\begin{center}
$g_{\omega\Lambda}$ = $g_{\omega\Sigma}$ = $2g_{\omega\Xi}$ = $\frac{2}{3}g_{\omega N}$
\end{center}
\begin{center}
 $2g_{\phi\Lambda}$ = $2g_{\phi\Sigma}$ = $g_{\phi\Xi}$ = $\frac{-2\sqrt{2}}{3} g_{\omega N}$
\end{center}

The scalar couplings are determined by fitting the hyperonic potential,
\begin{eqnarray}
U^{(N)}_Y = g_{\omega Y}\omega_0 + g_{\sigma Y}\sigma_0
\end{eqnarray}
where Y stands for the hyperon and $\sigma_0$, $\omega_0$ are the values of the scalar and vector meson fields
at saturation density~\cite{8}. The values of $U^{(N)}_Y$ are taken from the available hypernuclear data.
The best known hyperonic potential is that of $\Lambda$, having a value of about $U^{(N)}_\Lambda$ = -30 MeV~\cite{36a}.
In case of $\Sigma$ and $\Xi$ hyperons, the potential depths are not as clearly known as in the case of $\Lambda$.
However, analyses of laboratory experiments indicate that at
nuclear densities the $\Lambda$-nucleon potential is attractive but the $\Sigma^-$ -nucleon potential
is repulsive~\cite{potential}.
Therefore, we have varied both $U^{(N)}_\Sigma$ and $U^{(N)}_\Xi$ in the range of -40 MeV to +40 MeV to investigate the 
properties of neutron star matter.

 For neutron star matter, with baryons and charged
leptons, the $\beta$-equilibrium conditions are guaranteed with
the following relations between chemical potentials for different
particles:
 \begin{eqnarray}
  \mu_p &=& \mu_{\Sigma^{+}} = \mu_n - \mu_e \nonumber \\
 \mu_{\Lambda} &=& \mu_{\Sigma^{0}} = \mu_{\Xi^{0}} = \mu_n \nonumber\\
 \mu_{\Sigma^{-}} &=& \mu_{\Xi^{-}} = \mu_n+\mu_e \nonumber\\
 \mu_{\mu} &=& \mu_e 
\end{eqnarray}
and the charge neutrality condition is fulfilled by
 \begin{equation}
  n_p + n_{\Sigma^{+}} = n_e+n_{\mu^{-}}+n_{\Sigma^{-}}+n_{\Xi^{-}}
 \end{equation}
 where $n_i$ is the number density of the {\it i'th} particle. The effective chemical
potentials of baryons and leptons can be given by
\begin{equation}
 \mu_B = \sqrt{{k_{F}^{B}}^2+{m_{B}^{\ast^2}}}+g_{\omega B}\omega+g_{\rho B}\tau_{3B}\rho
\end{equation} 

\vspace {1cm}

 \begin{equation}
 \mu_l = \sqrt {{K_F^{l}}^2+m_l^2}
 \end{equation}
where $m_{B}^{\ast} = m_B-g_{\sigma B}\sigma$ is the baryon effective mass and $K_F^l$ is the Fermi momentum of the lepton (e, $\mu$).
 The EoS of neutron star matter can be given by,
 \begin{figure*}[t]
 \begin{minipage}[b]{.5\textwidth}%
\includegraphics[width = 2.0in,height = 3.7in, angle = 270]{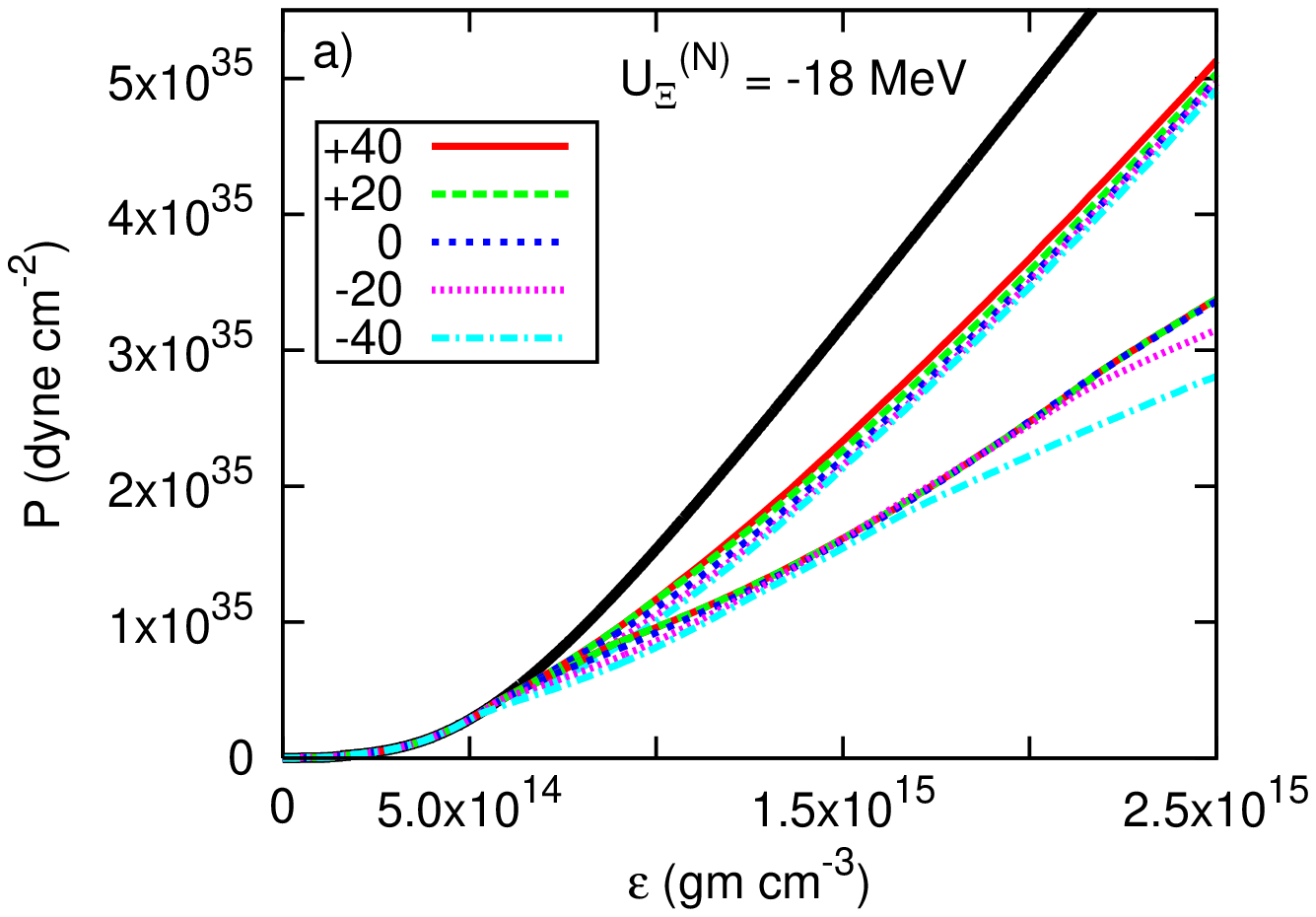}
\end{minipage}\hfill
\begin{minipage}[b]{.5\textwidth}
\includegraphics[width = 2.1in,height = 3.4in, angle = 270]{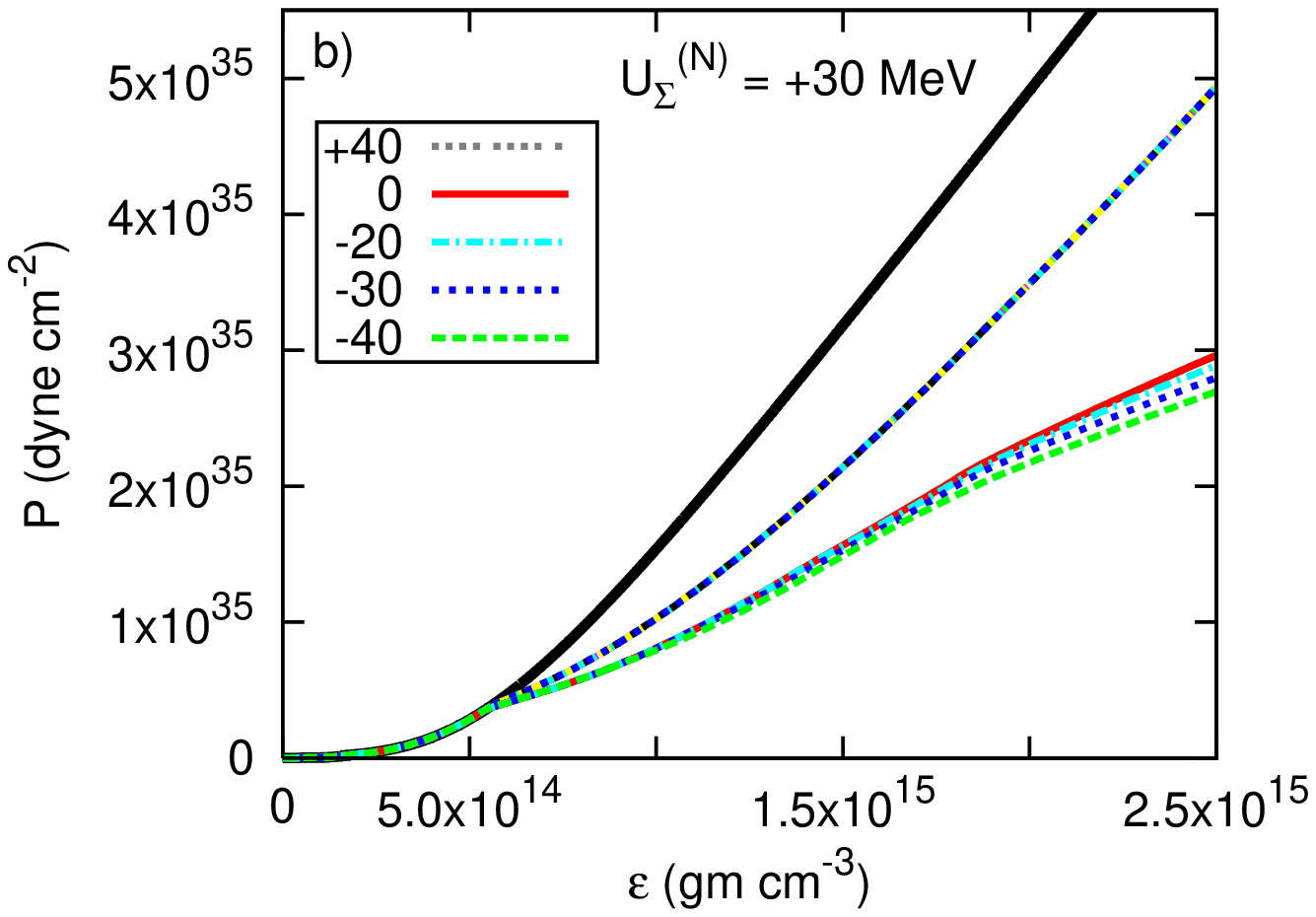}
\end{minipage}

\caption{ (color online) a) EoS obtained with varying  $U^{(N)}_\Sigma$ at fixed $U^{(N)}_\Xi$. 
 The upper branch shows the EoS for a system containing nucleons, leptons and all the non 
 strange mesons. The middle branch shows the EoS for a system containing the whole baryon octet, the leptons 
 and $\sigma$, $\omega$, $\rho$ and $\phi$ mesons. The lower branch shows the EoS
 for the particles contained in the middle branch except $\phi$. 
 b) EoS obtained with varying  $U^{(N)}_\Xi$ at fixed $U^{(N)}_\Sigma$. The compositions 
 of the upper, middle and lower branches are same as those of a) respectively. 
 \label{eos}}
\end{figure*}

\begin{widetext}
\begin{eqnarray}
{\varepsilon} &=& \frac{1}{2}m_\sigma^2 \sigma^2
+ \frac{\kappa}{6} g_{\sigma N}^3 \sigma^3 + \frac{\lambda}{24}  g_{\sigma N}^4 \sigma^4 + \frac{1}{2} m_\omega^2 \omega^2 
+ \frac{\zeta}{8} 
g_{\omega N}^4 \omega^4 
+ \frac{1}{2} m_\rho^2 \rho^2 + 3\Lambda_v g_{\rho N}^2 g_{\omega N}^2{\omega}^2{\rho}^2 \nn\\
&& + \frac{1}{2}m_\phi^2 \phi^2 + \sum_B\frac{\gamma_B}{(2\pi)^3} 
\int_0^{k_{F}^B} \sqrt{k^2+m^{* 2}_B} \ d^3 k 
 + \frac{1}{\pi^2}\sum_l\int_0^{K_F^l} \sqrt{k^2+m^2_l} \ k^2 dk
\end{eqnarray}

\begin{eqnarray}
P &=& - \frac{1}{2}m_\sigma^2 \sigma^2
- \frac{\kappa}{6} g_{\sigma N}^3 \sigma^3 - \frac{\lambda}{24} g_{\sigma N}^4  \sigma^4 + \frac{1}{2} m_\omega^2 \omega^2 +  \frac{\zeta}{24}
g_{\omega N}^4 \omega^4 + \Lambda_v g_{\rho N}^2 g_{\omega N}^2{\omega}^2{\rho}^2 
+ \frac{1}{2} m_\rho^2 \rho^2 \nn\\
&& + \frac{1}{2}m_\phi^2 \phi^2 + \frac{1}{3}\sum_B \frac{\gamma_B}{(2\pi)^3} 
\int_0^{k_F^B}\frac{k^2 \  d^3 k}{(k^2+m^{* 2}_B)^{1/2}}
 + \frac{1}{3} \sum_{l} \frac{1}{\pi^2}
\int_0^{K_F^l}\frac{k^4 \ dk}{(k^2+m^2_l)^{1/2}}~\nn\\
\label{pr}
\end {eqnarray}
\end{widetext}
where $\varepsilon$ and $P$ stand for energy density and pressure respectively and $\gamma_B$ is the baryon spin-isospin degeneracy factor.

 \begin{figure}[htb]
\hskip 1.5cm
 \begin{minipage}{.01in}
 \rotatebox{90}{\bf particle fractions}
 \end{minipage}%
 \begin{minipage}{\dimexpr\linewidth-2.50cm\relax}
 {\includegraphics[width = 2.8in,height = 1.7in]{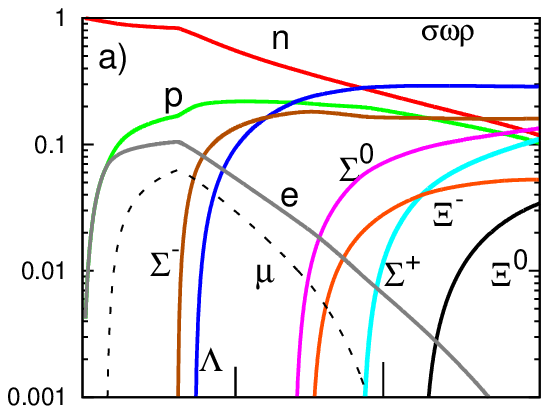}}\\
 \vspace{-6.6mm}\bigskip{}
 {\includegraphics[width = 2.8in, height = 1.7in]{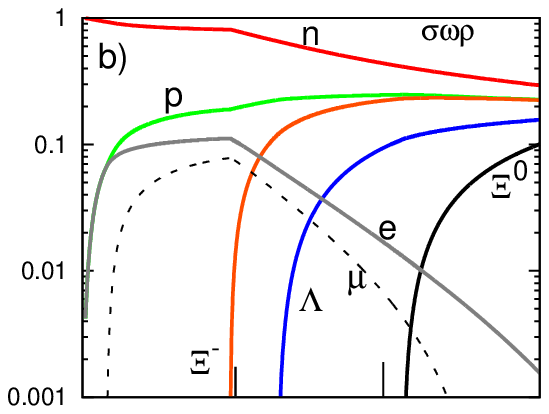}}\\
 \vspace*{-6.5mm}\bigskip{}
 {\includegraphics[width = 2.8in, height = 1.7in]{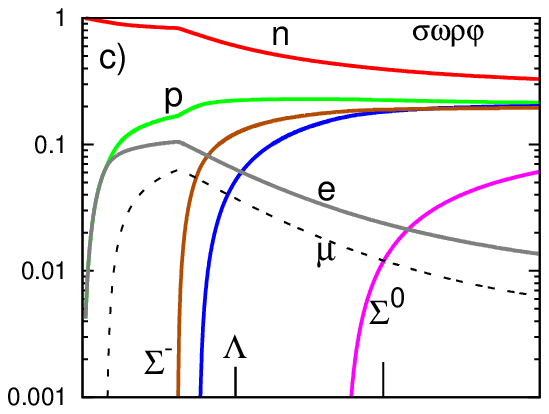}}\\
 \vspace*{-6.5mm}\bigskip{}
 {\includegraphics[width = 2.8in, height = 1.7in]{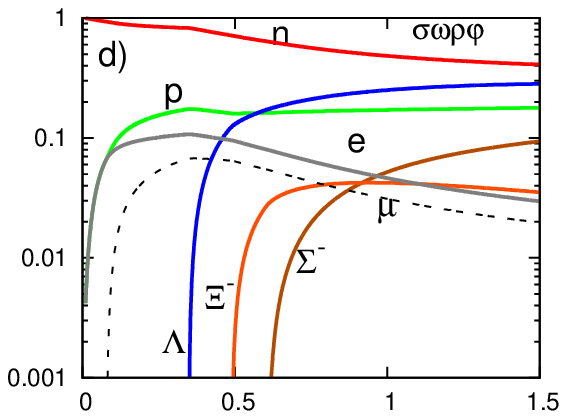}}\\
 \vspace*{0.1cm}\hspace*{2.0cm}{\bm {$n_B$ $(fm^{-3})$}}
 \end{minipage}
 
 \caption{ (color online) Particle fractions for different $\Sigma$ potential depths: a)
 for ``$\sigma\omega\rho$''
  with $U^{(N)}_\Sigma = -30$ MeV, b) for ``$\sigma\omega\rho$''
  with $U^{(N)}_\Sigma = +30$ MeV, c) for ``$\sigma\omega\rho\phi$''
  with $U^{(N)}_\Sigma = -30$ MeV, d) for ``$\sigma\omega\rho\phi$''
  with $U^{(N)}_\Sigma = +30$ MeV. $U^{(N)}_\Xi$ is fixed at -18 MeV in each case.\label{pf1}}
 \end{figure}

In fig. \ref{eos} we plot the EoS for different values of the hyperonic potentials.
 The upper branch is for the usual nuclear matter which
  does not contain any strange particle. 
The middle and lower branches 
are for full baryon octet, leptons and $\sigma$, $\omega$, $\rho$ mesons.  In addition, the middle branch contains the $\phi$ meson. 
 In the left 
panel, {\it i.e.} in fig. \ref{eos}a, we keep $U^{(N)}_\Xi$ fixed at -18 MeV,  
this value is generally adopted from hypernuclear experimental data~\cite{pot}. For the middle and lower branches 
 we vary the $\Sigma$ potential from -40 MeV to +40 MeV in steps of 20 MeV. The lower branch shows that for an attractive $\Sigma$ potential 
the EoS gets stiffer as $U_{\Sigma}^{(N)}$ increases. However as $U_{\Sigma}^{(N)}$ becomes 
positive the EoS seems to become independent of $U_{\Sigma}^{(N)}$. We see from fig. \ref{eos}a that for 
$U_{\Sigma}^{(N)} > 0$ MeV the EoS remains identical to that for $U_{\Sigma}^{(N)} = 0$ MeV.
However, once we add $\phi$ meson to the system, the EoS continues to get stiffer as $U_{\Sigma}^{(N)}$ moves to more positive side (middle branch of fig. \ref{eos}a). 

  \begin{figure}[t]
 \begin{minipage}{.01in}
 \rotatebox{90}{\bf particle fractions}
 \end{minipage}%
 \begin{minipage}{\dimexpr\linewidth-2.50cm\relax}
 {\includegraphics[width = 2.8in,height = 1.7in]{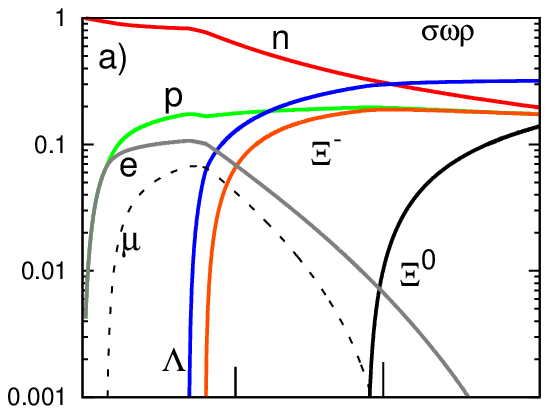}}\\
 \vspace*{-6.4mm}\bigskip{}
 {\includegraphics[width = 2.8in, height = 1.7in]{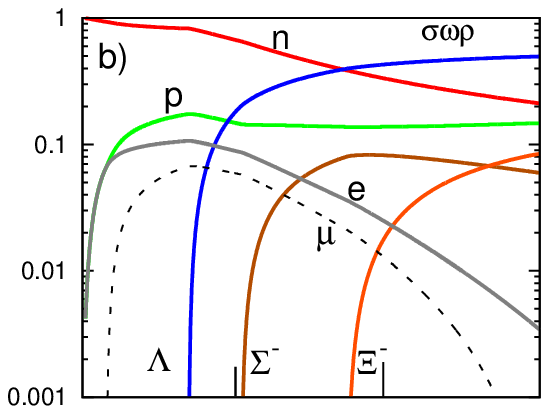}}\\
 \vspace*{-6.3mm}\bigskip{}
 {\includegraphics[width = 2.8in, height = 1.7in]{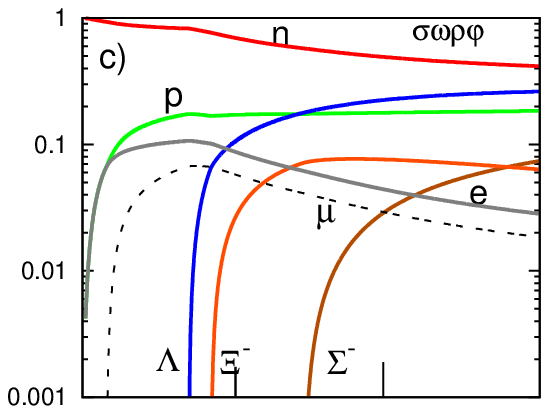}}\\
 \vspace*{-6.3mm}\bigskip{}
 {\includegraphics[width = 2.8in, height = 1.7in]{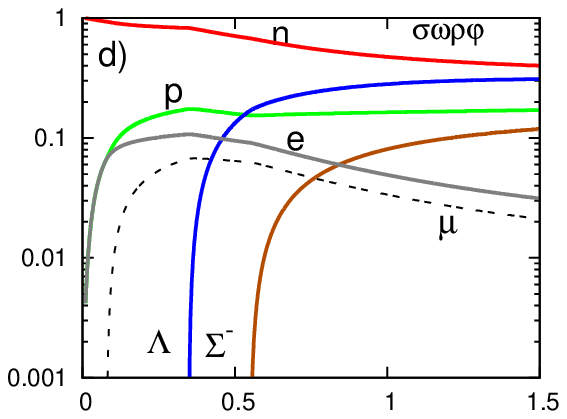}}\\
 \vspace*{0.1cm}\hspace*{2.0cm}{\bm {$n_B$ $(fm^{-3})$}}
 \end{minipage}
 
  \caption{ (color online) Particle fractions for different $\Xi$ potential depths: a) for ``$\sigma\omega\rho$''
  with $U^{(N)}_\Xi = -30$ MeV, b) for ``$\sigma\omega\rho$''
  with $U^{(N)}_\Xi = +30$ MeV, c) for ``$\sigma\omega\rho\phi$''
 with $U^{(N)}_\Xi = -30$ MeV, d) for ``$\sigma\omega\rho\phi$''
 with $U^{(N)}_\Xi = +30$ MeV. $U^{(N)}_\Sigma$ is fixed at +30 MeV in each case.\label{pf2}}
 \end{figure}

 We then fix $U^{(N)}_\Sigma$ and vary $U^{(N)}_\Xi$. 
 This is represented in fig. \ref{eos}b, where we have fixed the value of $U^{(N)}_\Sigma = +30$ MeV 
 (adopted from hypernuclear experimental data~\cite{pot}). We vary $U^{(N)}_\Xi$ from -40 MeV to +40 MeV. 
We see that for the lower branch, {\it i.e} the case without the $\phi$ meson, the EoS 
gets stiffer with the increase in $\Xi$ potential up to $U^{(N)}_\Xi = 0$ MeV. 
However, for positive values of $U^{(N)}_\Xi$ the EoS remains unchanged. 
Adding an extra repulsion to the system by including the $\phi$ meson changes
the scenario altogether. The EoS becomes totally independent of the $\Xi$ potential (middle branch of fig. \ref{eos}b). 
From figures 1a and 1b one can generally conclude that the inclusion 
of $\phi$ meson makes the EoS stiffer, however, hyperonic EoS is much softer than the usual 
nuclear matter EoS. 

\begin{figure*} [t]
\begin{minipage}[t]{.5\textwidth}
\includegraphics[width = 2.0in,height = 3.8in, angle = 270]{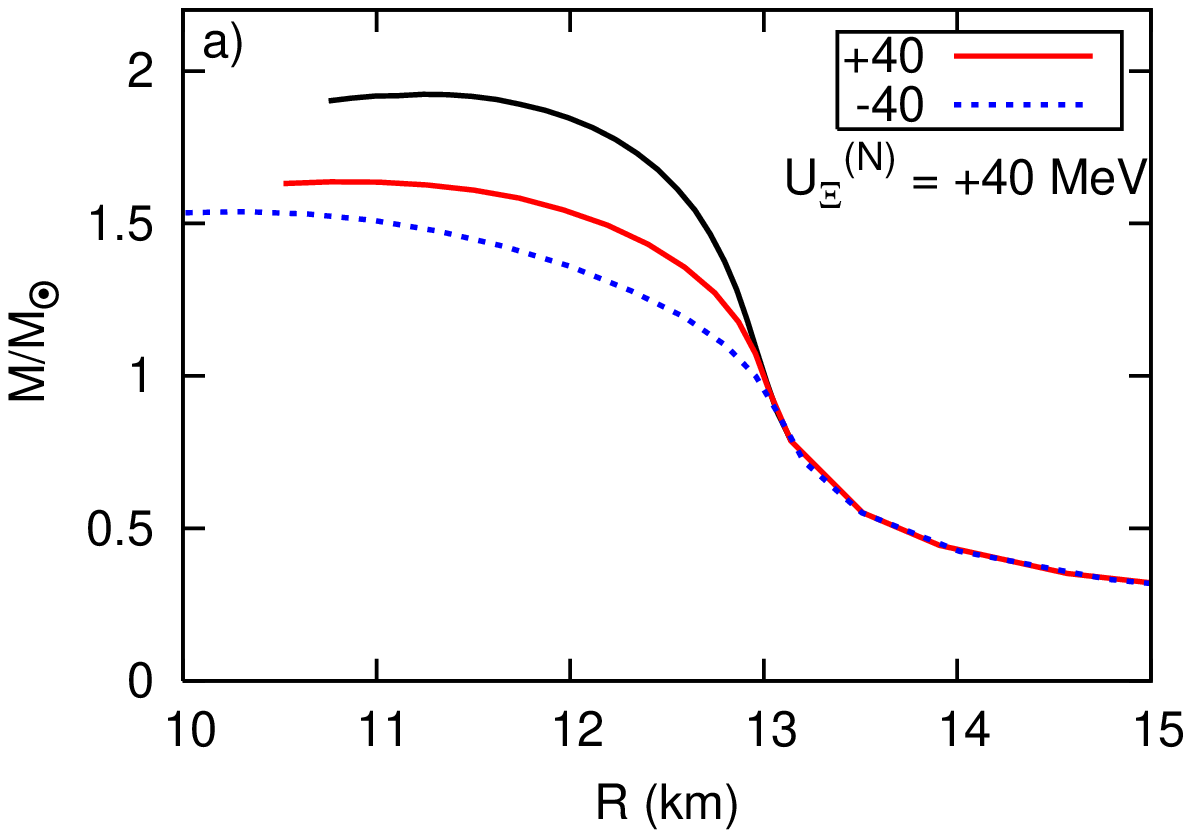}
\end{minipage}\hfill
\begin{minipage}[t]{.5\textwidth}
\includegraphics[width = 2.0in,height = 3.8in, angle = 270]{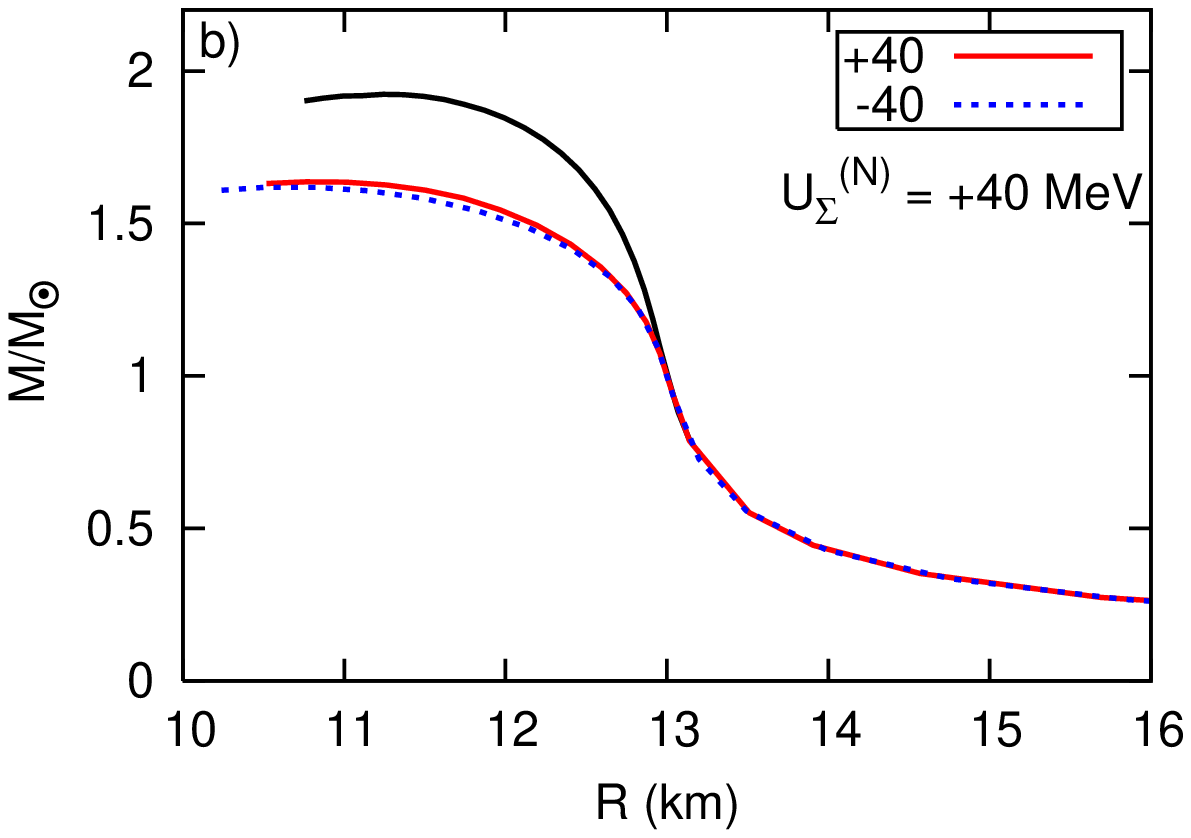}
\end{minipage}

\caption{ (color online) Mass-radius curves for static star fixing the a) $\Xi$ potential depth 
at $U^{(N)}_\Xi = +40$ MeV and varying the $U^{(N)}_\Sigma$.
b) $\Sigma$ potential depth 
at $U^{(N)}_\Sigma = +40$ MeV and varying the $U^{(N)}_\Xi$. The uppermost curve in each case corresponds to the pure nuclear matter.}
\label{static}
\end{figure*}

In fig. \ref{pf1} we have plotted the particle fractions for an attractive 
 $\Sigma$ potential $U^{(N)}_\Sigma = -30$ MeV and a repulsive potential $U^{(N)}_\Sigma = +30$ MeV keeping $U^{(N)}_\Xi$ fixed 
 at -18 MeV, with and without $\phi$ in each case.
From fig. \ref{pf1}a, when $\phi$ is not present, we see that all the hyperons contribute 
to the particle fractions for an
attractive $\Sigma$ potential whereas for repulsive $U^{(N)}_\Sigma$ there is no $\Sigma$
present in the matter (fig. \ref{pf1}b). The appearance of $\Lambda$ is also pushed to higher 
density compared to the case of an attractive potential. 
When $\phi$ is included in the system
 $\Sigma^0$ and $\Sigma^-$ appear with $\Lambda$ for $U^{(N)}_\Sigma = -30$ MeV (fig. \ref{pf1}c). However, for $U^{(N)}_\Sigma = +30$ MeV
(fig. \ref{pf1}d), the threshold of $\Sigma^-$ is pushed to higher density compared to the case of $U^{(N)}_\Sigma = -30$ MeV, $\Sigma^0$ disappears
and $\Xi^-$ appears in the system.
We also note that in the case of attractive $\Sigma$ potential, $\Sigma^-$ is always the first hyperon to appear in the system.
For repulsive $U^{(N)}_\Sigma$, $\Xi^-$ appears before others in the ``$\sigma\omega\rho$'' case
and $\Lambda$ is the the first hyperon to appear in case of ``$\sigma\omega\rho\phi$''.

 From fig. \ref{pf1} we see that for negative values of  $U^{(N)}_\Sigma$, the $\Sigma$'s are
bound in matter and the effective mesonic interaction would be more attractive as the potential gets deeper. 
As a result, the EoS gets softer with more attractive $U^{(N)}_\Sigma$ (see fig. \ref{eos}a). For 
$U^{(N)}_\Sigma\geq 0$, $\Sigma$'s are no longer bound to matter and the effective
mesonic interaction becomes more and more repulsive with increasing
$U^{(N)}_\Sigma$. This should, in principle, stiffen the EoS. However, for the
``$\sigma\omega\rho$'' case, up to neutron star
densities, {\it i.e} about $n_B \lesssim (4-7)n_0$, $\Sigma$'s are not present in the 
matter when the potential is repulsive and hence the EoS up to these densities
becomes insensitive to $U^{(N)}_\Sigma$.

\begin{figure*}[t]
\begin{minipage}[t]{.5\textwidth}
\includegraphics[width = 2.0in,height = 3.8in, angle = 270]{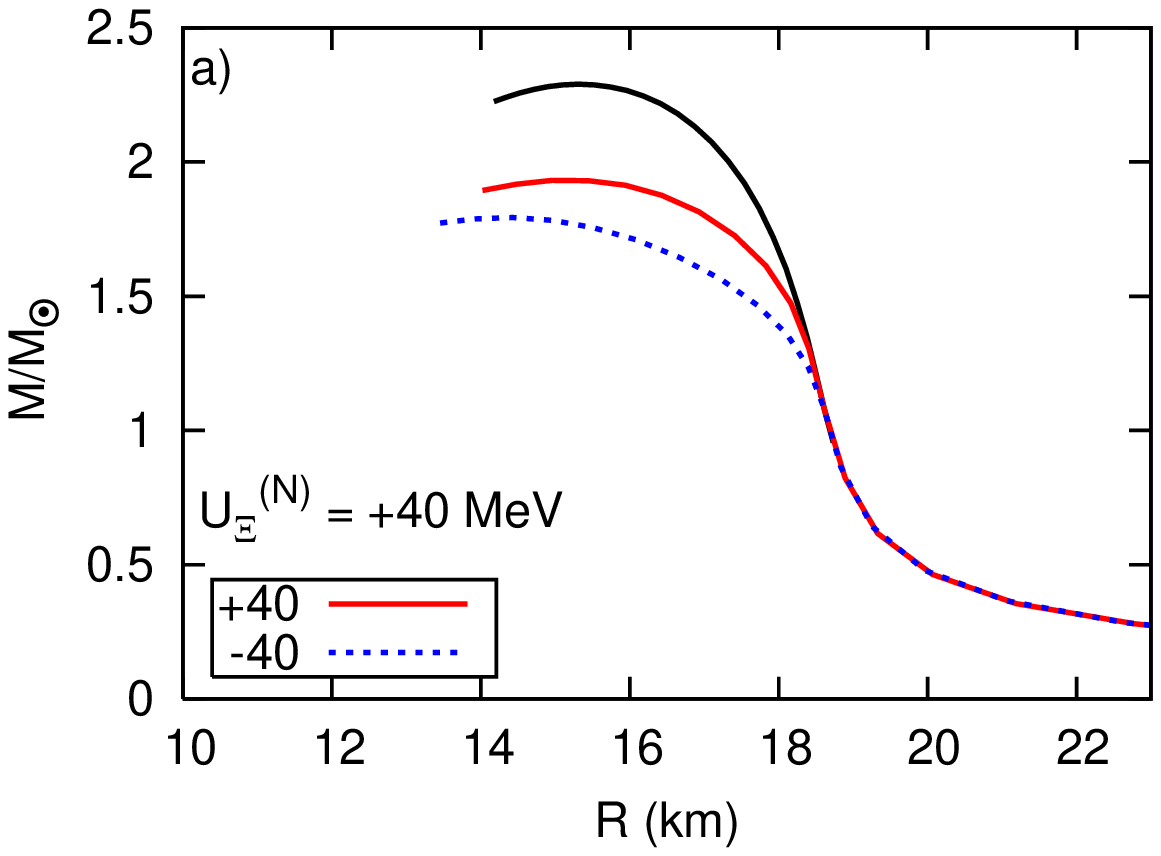}
\end{minipage}\hfill
\begin{minipage}[t]{.5\textwidth}
\includegraphics[width = 2.0in,height = 3.8in, angle = 270]{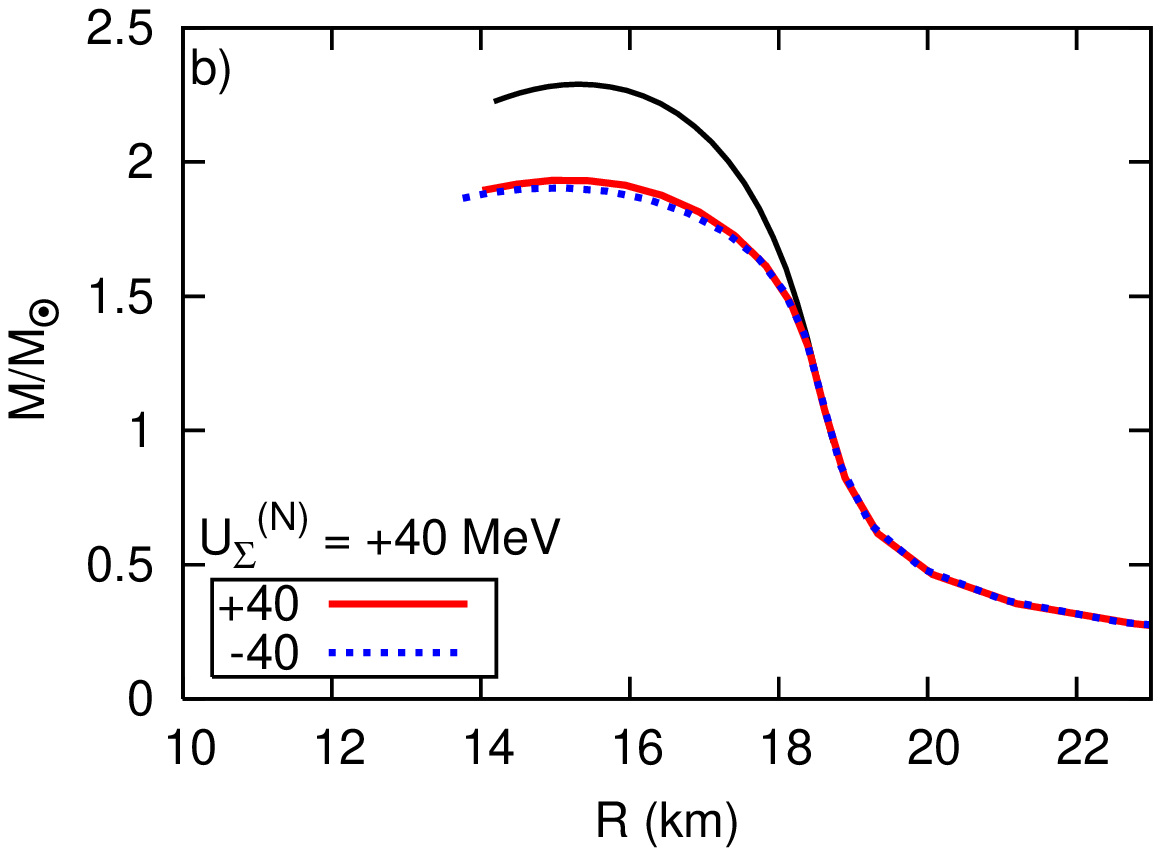}
\end{minipage}
\caption{ (color online) Mass-radius curves for rotating stars for two cases: a) $U^{(N)}_\Xi = +40$ MeV
and $-40 MeV \leq U^{(N)}_\Sigma \leq +40$ MeV and b) $U^{(N)}_\Sigma = +40$ MeV
and $-40 MeV \leq U^{(N)}_\Xi \leq +40$ MeV. The uppermost curve in each case corresponds to the pure nuclear matter.\label{MR3}}
\end{figure*}

In fig. \ref{pf2} the particle fractions are plotted for an attractive $\Xi$ potential $U^{(N)}_\Xi = -30$ MeV 
and a repulsive potential $U^{(N)}_\Xi = +30$ MeV keeping $U^{(N)}_\Sigma$ fixed at +30 MeV.
We see that in the first case {\it i.e.} when $\phi$ is not present and the potential is attractive 
 (fig. \ref{pf2}a), all the hyperons except $\Sigma$'s are present in the system
and the $\Lambda$ hyperon dominates. When the $\Xi$ potential becomes positive (fig. \ref{pf2}b) $\Xi^0$
disappears and the threshold for appearance of $\Xi^-$ shifts to much higher density. However $\Sigma^-$ is present in matter in this
potential and it appears before $\Xi^-$. When $\phi$ is introduced in the system, for an attractive $\Xi$ potential (fig. \ref{pf2}c), again $\Sigma^-$ and $\Xi^-$ 
are present along with $\Lambda$. However, the difference from fig. \ref{pf2}b {\it i.e} ``$\sigma\omega\rho$'' case
and $U^{(N)}_\Xi\geq$0 is that, here $\Xi^-$ appears much before $\Sigma^-$.
In the last case (fig. \ref{pf2}d), we see that as a result of the combined effects of inclusion of $\phi$ and repulsive potentials,
only the $\Lambda$ and $\Sigma^-$ are present in the system.
 From both figures \ref{pf1} and \ref{pf2}, we see that, inclusion of 
 $\phi$ meson decreases the density of hyperons. Since $\phi$ is a strange particle, further strangeness is suppressed and 
 as a result the hyperon densities are reduced compared to the ``$\sigma\omega\rho$'' case.

\section {static and rotating stars}

In this section we are going to discuss the properties of static and rotating axisymmetric stars using the EoS
which we have studied in the last section. The EoS without $\phi$ meson is softer compared 
to that with $\phi$ meson. So we do not discuss the EoS without $\phi$ as it results in less maximum mass. 


The stationary, axisymmetric space-time used to model the compact stars are defined through the metric
\begin{eqnarray}
 ds^2 = -e^{\gamma+\rho} dt^2 + e^ {2\alpha}(dr^2+r^2d\theta^2)\nn\\
 + e^{\gamma-\rho}r^2 sin^2{\theta}(d\phi-\omega dt)^2 
\end{eqnarray}
\noindent where $\alpha$, $\gamma$ , $\rho$ and $\omega$ are the gravitational potentials which
depend on r and $\theta$ only.

In this work we adopt the procedure of Komatsu \etal~\cite{37} to look into the observable properties of static and rotating
stars. Einstein's equations for the three gravitational potentials $\gamma$, $\rho$ and $\omega$ can be solved using Green's 
function technique. The fourth potential $\alpha$ can be determined using these three potentials. Once these potentials are 
determined one can calculate all the observable quantities using those. The solution of the potentials and
hence the determination of physical quantities is numerically quite an involved process. For this purpose the 
``rns'' code~\cite{39} is used in this work. This code, developed by Stergoilas, is very efficient in calculating the rotating star 
observables.

We discuss the properties of static stars first. In fig. \ref{static} we have plotted the mass-radius curves 
of static stars 
using the EoS with ``$\sigma\omega\rho\phi$''. A plot for the pure nuclear matter
case is also given for comparison (uppermost curve of both the panels). The maximum mass of pure nuclear matter 
star in the static case is $1.92 M_\odot$ with a radius of $11.24$ km. We have found that the mass of hyperonic star
becomes maximum for $U_\Sigma^{N} = +40$ MeV and $U_\Xi^{N} \geq 0$ MeV. Hence in fig. \ref{static} and
fig. \ref{MR3} we have shown the effect of these potentials on the maximum mass of neutron stars by fixing one of the 
potentials at +40 MeV and varying the other. The left panel, {\it i.e.} 
fig. \ref{static}a, corresponds to $U_\Xi = +40$ MeV and $U_{\Sigma}$ varying from -40 MeV to +40 MeV. In 
the right panel, {\it i.e.} in fig. \ref{static}b,  it is the other way round. 
From fig. \ref{static}a one can see that the maximum mass of the star increases with $U_\Sigma^{(N)}$. 
For $U_\Sigma^{(N)} = +40$ MeV the maximum mass is $1.62 M_\odot$ with a radius of $10.82$ km. The central 
energy density of such a star is $\epsilon_c = 2.46 \times 10^{15} gm \,\, cm^{-3}$.  This is a reflection of the 
EoS shown in fig. \ref{eos}a, which shows that the EoS becomes stiffer with increase in $U_\Sigma^{(N)}$. However, 
as seen from fig. \ref{static}b, the maximum mass of static stars is insensitive 
to $U_\Xi^{(N)}$, which should be obvious from fig. \ref{eos}b as the EoS is independent of the cascade potential.
Furthermore, from fig. \ref{pf2}d one can see that there is no cascade present in the medium. So the 
insensitivity of the EoS and hence the maximum mass, towards the cascade potential is expected. One should note that the maximum 
mass we obtain for the static stars is less than the observed mass of PSR J$0348+0432$ . So the static stars with hyperons 
in the IUFSU 
parameter set can not incorporate a maximum mass $\sim 2M_\odot$. This result is consistent with the findings in Ref.~\cite{Agrawal}.
However, since both of the observed
$\sim 2M_\odot$ stars are pulsars, it would be
a better idea to compare the observations with results from the rotating stars, which we do in the next part.

\begin{figure}[h]
\resizebox{8cm}{!}{
 \includegraphics{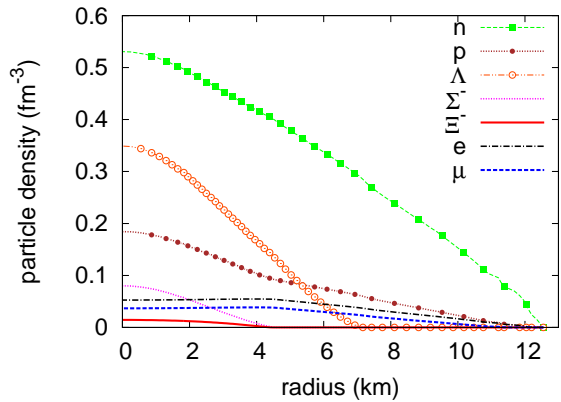}}
\caption{ (color online) Particle densities varying with radius along the equator.
The potential depths for which particle densities are plotted are $U^{(N)}_\Xi = 0$ and $U^{(N)}_\Sigma = +40$ MeV.
 \label{pd}}
\end{figure}

In fig. \ref{MR3} we plot the mass-radius curves for stars rotating with Keplerian velocities, for two cases. In 
fig. \ref{MR3}a we fix the cascade potential 
at $U^{(N)}_\Xi = +40$MeV and vary $U^{(N)}_\Sigma$ from $-40$MeV to $+40$MeV. In fig. \ref{MR3}b it 
is the other way round. The pure nuclear matter case is also shown in the uppermost curve. The maximum mass for the pure 
nucleonic star is $2.29 M_\odot$ with a radius of $15.31$ km. We see that
the maximum mass obtained for a rotating star with hyperonic core is $1.93M_{\odot}$ with a radius of $14.7$ km in
the Keplerian limit with angular velocity $\Omega = 0.86\times 10^4 s^{-1}$,
 for $U^{(N)}_\Sigma = +40$ MeV and 
$U^{(N)}_\Xi \geq 0$. 
As in the case of static sequence, we see that the maximum mass for the rotating case also increases with $U^{(N)}_\Sigma$ as we go 
towards more positive values of this potential. At $U^{(N)}_\Sigma = -40$ MeV we get a maximum mass of $1.79 M_{\odot}$
whereas for $U^{(N)}_\Sigma = +40$ MeV the maximum mass is $1.93 M_{\odot}$. 
The effect of $U^{(N)}_\Xi$ is much less significant on the maximum mass.
From $U^{(N)}_\Xi = -40$ MeV 
to $U^{(N)}_\Xi = +40$ MeV mass is changed only by $\bigtriangleup M = 0.03 M_{\odot}$. 

In order to have a look at the composition of the maximum mass star, we have plotted the particle densities as a function 
of radius along the equator in fig. \ref{pd}. For $U^{(N)}_\Xi = 0$ and $U^{(N)}_\Sigma = +40$ MeV, 
 we see that a fair amount of hyperons are present in the core. There are $\Lambda$, $\Sigma^-$ and $\Xi^-$
 present. Another interesting observation is that near the core, 
the density of $\Lambda$ is much more compared to that of protons and it continues up to a distance of about 5 km from the center.

\section{Summary and conclusions}

To summarize, we have studied the static and rotating axisymmetric stars with hyperons using IUFSU model. The original FSUGold 
parameter set has been very successful in describing the properties of finite nuclei. With the discovery of highly massive neutron stars 
the reliability of this model was questioned. It was then revised in the form of IUFSU to accommodate such highly massive stars
leaving the low density finite nuclear properties unchanged. In this
work we have studied this new parameter set in the context of the possibility of having a hyperonic core in such massive stars.

We have included the full octet of baryons in IUFSU. The EoS gets softened due to the inclusion of hyperons
whereas the inclusion of the $\phi$ meson makes the EoS stiffer. We have also investigated the influence of 
$\Sigma$ and $\Xi$ potentials on the EoS.

 For static stars with hyperonic core we get a maximum mass of 
 $1.62 M_\odot$. So IUFSU with hyperons cannot reproduce the observed 
mass of static stars. However, as the observed $\sim 2M_\odot$ neutron stars are both pulsars, we 
compare the results in the rotating limit. In the Keplerian limit we get a maximum mass of $1.93 M_{\odot}$, which is within the
3$\sigma$ limit of the mass of PSR J$0348+0432$ and 1$\sigma$ limit of the earlier observation of PSR J$1614-2230$. 
We have looked at the particle densities inside the star 
having the maximum mass and found that a considerable amount of hyperons are present near the core.
Therefore, our results are consistent with the recent observations of highly massive pulsars confirming the presence of hyperons
in the core of such massive neutron stars.

To conclude, IUFSU model, which reproduces the properties of finite nuclei quite successfully also reproduces the recent observations 
of $\sim 2M_{\odot}$ stars, in case of stars having exotic core and rotating in the Keplerian limit. It will be interesting to see 
whether such a star can hold a quark core. Related work is in progress. 

\section{Acknowledgement}
This work is funded by the University Grants Commission (RFSMS, DSKPDF and DRS) and  Department of Science and Technology, 
Government  of India.


\begin{thebibliography}{99}
\bibitem{science}J. Antoniadis {\it et al.} {\it Science} {\bf 340}, (2013) 6131.
\bibitem{Nature}P. B. Demorest, T. Pennucci, S. M. Ransom, M. S. E. Roberts and J. W. T. Hessels, {\it Nature} {\bf 467}, (2010) 1081.
\bibitem{2}E. Massot, J. Margueron1 and G. Chanfray, {\it \epl} {\bf 97}, (2012) 39002.
\bibitem{3}M. Baldo, G. F. Burgio and H.-J. Schulze, {\it Phys. Rev. C} {\bf 61}, (2000) 055801.
\bibitem{4}I. Vidana, A. Polls, A. Ramos, L. Engvik and M. Hjorth-Jensen, {\it Phys. Rev. C} {\bf 62}, (2000) 035801.
\bibitem{5}H. \DJ{}apo, B.-J. Schaefer and J. Wambach, {\it Phys. Rev. C} {\bf 81}, (2010) 035803.
\bibitem{6}J. M. Lattimer and M. Prakash, {\it From Nuclei to Stars: Festschrift in Honor of Gerald Brown,} p.{\bf 275}, World Scientific,
Singapore, (2011).
\bibitem{7}N. K. Glendenning {\it Astrophys. J.} {\bf 293}, (1985) 470. 
\bibitem{8}N. K. Glendenning and S. A. Moszkowski, {\it Phys. Rev. Lett.} {\bf 67}, (1991) 2414.
\bibitem{9}R. Knorren, M. Prakash and P. J. Ellis, {\it Phys. Rev. C} {\bf 52}, (1995) 3470.
\bibitem{10}S. Balberg and A. Gal, {\it Nucl.Phys. A} {\bf 625}, (1997) 435.
\bibitem{11}S. Pal, M. Hanauske, I. Zakout, H. St$\ddot{o}$ecker and W. Greiner, {\it Phys. Rev. C} {\bf 60}, (1999) 015802.
\bibitem{12}M. Hanauske, D. Zschiesche, S. Pal, S. Schramm, H. St$\ddot{o}$ecker and W. Greiner,  {\it Astrophys. J.} {\bf 537}, (2000) 958.
\bibitem{13}S. Schramm and D. Zschiesche, {\it J. Phys. G} {\bf 29}, (2003) 531.
\bibitem{14}W. H. Long, B. Y. Sun, K. Hagino and H. Sagawa, {\it  Phys. Rev. C} {\bf 85}, (2012) 025806.
\bibitem{15}H. Huber, M. K. Weigel and F. Weber, Z. {\it Naturforsch.} {\bf 54A}, (1999) 77.
\bibitem{16}F. Hofmann, C. M. Keil and H. Lenske, {\it Phys. Rev. C} {\bf 64}, (2001) 034314.
\bibitem{17}J. Rikovska-Stone, P. Guichon, H. Matevosyan and A. Thomas, {\it Nucl. Phys. A} {\bf 792}, (2007) 341.
\bibitem{18}S. K. Dhiman, R. Kumar and B. K. Agrawal, {\it Phys. Rev. C} {\bf 76}, (2007) 045801.
\bibitem{19}V. Dexheimer and S. Schramm, {\it Astrophys. J.} {\bf 683}, (2008) 943.
\bibitem{20}I. Bombaci, P. K. Panda, C. Providencia and I. Vidana, {\it Phys. Rev. D} {\bf 77}, (2008) 083002.
\bibitem{21}R. Cavagnoli, D. P. Menezes and C. Providencia, {\it Phys. Rev. C} {\bf 84}, (2011) 065810.
\bibitem{22}M. Baldo, G. F. Burgio and H.-J. Schulze, {\it Phys. Rev. C} {\bf 58}, (1998) 3688.
\bibitem{23}S. Nishizaki, T. Takatsuka and Y. Yamamoto, {\it Prog. Theor. Phys.} {\bf 108}, (2002) 703.
\bibitem{24}H.-J. Schulze, A. Polls, A. Ramos and I. Vidana, {\it Phys. Rev. C} {\bf 73}, (2006) 058801.
\bibitem{25}H.-J. Schulze and T. Rijken, {\it Phys. Rev. C} {\bf 84}, (2011) 035801.
\bibitem{26}D. Logoteta, I. Vidana, C. Providencia, A. Polls and I. Bombaci, {\it J. Phys.: Conf. Ser.} {\bf 342}, (2012) 012006.
\bibitem{27}I. Bednarek, P. Haensel, J. L. Zdunik, M. Bejger and R. Ma\'{n}ka, {\it Astron. Astrophys.} {\bf 543}, (2012) A157.
\bibitem{28} R. Lastowiecki, D. Blaschke, H. Grigorian and S. Typel, {\it Acta Phys. Pol. B Proc. Suppl} {\bf 5}, (2012) 535.
\bibitem{29} A. R. Taurines, C. A. Z. Vasconcellos, M. Malheiro and M. Chiapparini, {\it Mod. Phys. Lett. A} {\bf 15}, (2000) 1789.
\bibitem{30} L. Bonanno and A. Sedrakian, {\it Astron. Astrophys.} {\bf 539}, (2012) A16.
\bibitem{Gupta} N. Gupta and P. Arumugam, {\it \PR C} {\bf 85}, (2012) 015804.
\bibitem{Agrawal1}B. K. Agrawal, A. Sulaksono and P. -G. Reinhard, {\it \NP A} {\bf 882}, (2012) 1.
\bibitem{30a} S. Weissenborn, D. Chatterjee and J. Schaffner-Bielich, {\it Phys. Rev. C} {\bf 85}, (2012) 065802.
\bibitem{30b} B. G. Todd-Rutel and J. Piekarewicz {\it Phys. Rev. Lett.} {\bf 95}, (2005) 122501.
\bibitem{31}  C. Wu and Z. Ren, {\it \PR C} {\bf 83} (2011) 025805.
\bibitem{32}F. J. Fattoyev, C. J. Horowitz, J. Piekarewicz and G. Shen, {\it \prc} {\bf 82}, (2010) 055803.
\bibitem{33} G. A. Lalazissis, J. K$\ddot{o}$nig and P. Ring, {\it Phy. Rev.} {\bf C 55}, (1997) 540. 
\bibitem{34} C. B. Dover and A. Gal, {\it Prog. Part. Nucl. Phys.} {\bf 12}, (1985) 171. 
\bibitem{35} J. Schaffner, C. B. Dover, A. Gal, C. Greiner, D. J. Millener and H. St$\ddot{o}$ecker, 
{\it Annals of Physics} {\bf 235}, (1994) 35. 
\bibitem{36a}D. J. Millener, C. B. Dover and A. Gal, {\it Phys. Rev. C} {\bf 38}, (1988) 2700;
 J. Schaffner, H. St$\ddot{o}$ecker and C. Greiner, {\it Phys. Rev. C} {\bf 46}, (1992) 322. 
\bibitem{potential}J. Mares, W. Friedman, A. Gal and B. K. Jennings, {\it Nucl. Phys. A} {\bf 594}, (1995) 311.
\bibitem{pot} J. Schaffner-Bielich and A. Gal, {\it \PR C} {\bf 62}, (2000) 034311.
\bibitem{37} H. Komatsu, Y. Eriguchi and  I. Hachisu , {\it Monthly Notices of Royal Astronomical Society} {\bf 237}, (1989) 355.
\bibitem{39} N. Stergioulas and J. H. Friedman, {\it Astrophys. J.} {\bf 444}, (1995) 306.
\bibitem{Agrawal} B. K. Agrawal, A. Sulaksono, P. -G. Reinhard, {\it Nucl. Phys. A}{\bf 882} (2012) 1.
\end{thebibliography}
\end{document}